\pdfoutput=1
\documentclass[pra,twocolumn,showpacs,groupedaddress,superscriptaddress,nofootinbib,floatfix,preprintnumbers,longbibliography]{revtex4-1}
\usepackage{amssymb,amsmath,graphicx,color}
\usepackage{physics}
\usepackage{bm}
\usepackage{dsfont}
\usepackage{appendix}
\usepackage{float}
\usepackage[tight]{subfigure}
\usepackage[export]{adjustbox}
\usepackage{braket}
\usepackage[colorlinks=true,citecolor=blue,linkcolor=red]{hyperref}
\usepackage{cleveref}
\usepackage{multirow}
\usepackage{rotating,booktabs}
\usepackage[verbose]{placeins}
\usepackage{color}
\newcommand\bea{\begin{eqnarray}}
\newcommand\eea{\end{eqnarray}}
\newcommand\barone{\underline{1}}

\newcommand{\ketsu}{\vert n_P,n_Q;\nu_{\underline{1}},\nu_0,\nu_1 \rangle}
\begin{document}
\title{Protecting gauge symmetries in the the dynamics of SU(3) lattice gauge theories}

\author{Emil Mathew}
\email{p20210036@goa.bits-pilani.ac.in}
\author{Indrakshi Raychowdhury}
\email{indrakshir@goa.bits-pilani.ac.in}
\affiliation{Department of Physics, BITS-Pilani,
K K Birla Goa Campus, Zuarinagar, Goa 403726, India}\affiliation{Center for Research in Quantum Information and Technology, Birla Institute of Technology and Science Pilani, Zuarinagar, Goa 403726, India}
\date{\today}
\begin{abstract}
Quantum simulation of the dynamics of a lattice gauge theory demands imposing on-site constraints. Ideally, the dynamics remain confined within the physical Hilbert space, where all the states satisfy those constraints. For non-Abelian gauge theories, implementing local constraints is non-trivial, as is keeping the dynamics confined in the physical Hilbert space, considering the erroneous quantum devices. SU(3) gauge group, albeit crucial for describing the strong interaction of nature, is notorious for studying via Hamiltonian simulation. 
This work presents a couple of symmetry protection protocols for simulating the exact dynamics of SU(3) gauge theory in $1+1$ dimension. The first protocol doesn't require imposing any local symmetry but relies on protecting global symmetries, which are Abelian with a preferred choice of framework, namely the loop-string-hadron framework. Generalization to a higher dimension is possible, however, the protection scheme needs to be local for that case but is still Abelian and thus advantageous. The symmetry protection schemes presented here are important steps towards quantum simulating the full theory of quantum chromodynamics. 
\end{abstract}
\maketitle
\tableofcontents

\section{Introduction}

\noindent
The Hamiltonian formulation, first put forth by Kogut and Susskind \cite{Kogut:1974ag}, offers a sign-problem-free approach to studying the real-time dynamics of the gauge theories, which provides a promising framework for computation in the upcoming quantum computation era. 

The Hamiltonian obtained via temporal gauge fixing is defined on the spatial lattice while the time remains continuous. However, the Hamiltonian dynamics for gauge theory is constrained by a set of local constraints known as Gauss law. The generators of the Gauss law constraints are mutually non-commuting for non-Abelian gauge theories such as SU(2) or SU(3) gauge theories, the later one being the underlying theory of the strong interaction of nature are of crucial physical interest. The primary ingredients for SU(N) lattice gauge theories are a Hamiltonian $\hat H$ and constraints $\hat G^a(r)$, which satisfy:
\begin{eqnarray}
\left[ H, G^a(r) \right]=0, ~~\left[ G^a(r), G^b(r') \right]=i{f^{ab}}_c G^c(r) \delta_{r,r'}
\end{eqnarray}
$\forall r$  and $a=1,2,...,N^2-1$. The physical Hilbert space for a gauge theory is generally chosen as the space of states, which are annihilated by all the Gauss' law operators $G^a(r)$ that generate gauge transformations locally at each site $r$. In principle, one can choose other gauge symmetry sectors (characterized by an integer quantum number of Gauss law generator) as physical Hilbert space, yet the Hamiltonian dynamics remains confined to that sector. The conventional choice of gauge invariant sector leads to Wilson loops, strings, and hadrons as the physical degrees of freedom for the gauge theory \cite{gambini2000loops, Raychowdhury:2019iki, Mathur:2007nu}.

Performing Hamiltonian simulation in the gauge theory Hilbert space (without imposing Gauss law constraint) involves huge computational costs being wasted in gauge redundancy and is not at all suitable for quantum simulation of the same as quantum resources are even more expensive and limited. Hamiltonian simulation of gauge theories primarily demands (i) preparation of the initial state in the gauge invariant Hilbert space/ physical Hilbert space (ii) and guarantee protection of gauge invariance throughout the simulated dynamics. Both of these tasks stand difficult in context to simulate the theory using the conventional framework and on a quantum hardware, respectively. 

The conventional framework for Hamiltonian lattice gauge theory is given by the well-studied Kogut-Susskind \cite{Kogut:1974ag} formulation. Construction of a gauge invariant Hilbert space in the Kogut-Susskind formulation is possible but expensive \cite{davoudi2021search}. 
Another major obstacle when working with the gauge invariant Hilbert space is that it involves non-locality. Even in the $1+1$ dimension, one can effectively eliminate all the gauge degrees of freedom for a gauge theory at the cost of introducing non-local interaction among on-site fermionic degrees of freedom. Over the last decade, there has been a renewed interest in Hamiltonian simulations of lattice gauge theories \cite{cirac2012goals,dalmonte2016lattice,banuls2020simulating,aidelsburger2022cold,zohar2022quantum,bauer2023quantum}. There have been a series of work on addressing this issue in the context of low-dimensional theories, and towards developing novel frameworks particularly 1+1 dimensions, \cite{chandrasekharan1997quantum,Brower:1997ha,zohar2011confinement,Zohar:2014qma,Zohar:2015hwa,zohar2017digital,Zohar:2018cwb,Zohar:2019ygc,ciavarella2021trailhead,banuls2017efficient,kaplan2020gauss,unmuth2019gauge,bender2020gauge,Haase2021resourceefficient,bauer2023efficient,ashkenazi2022duality,zache2018quantum,buser2021quantum,ji2020gluon,pardo2023resource,stryker2021shearing}. On the same note, there have been a multitude of research done into applying one or more of these formalism to describe digital as well as analog simulation protocols in the current NISQ-era \cite{zohar2011confinement,banerjee2012atomic,Banerjee:2012xg,Zohar:2012xf,tagliacozzo2013simulation,Zohar:2015hwa,Tagliacozzo:2012vg,Mazza_2012,kasper2016schwinger,Gonzalez-Cuadra:2017lvz,zohar2017digital,Kasper_2017,Muschik_2017,lamm2019general,shaw2020quantum,armon2021photon,carena2022improved,davoudi2020towards,davoudi2021toward,klco2018quantum,klco20202,raychowdhury2020solving,dasgupta2022cold,paulson2021simulating,riechert2022engineering,Zhou_2022,homeier2023realistic,muller2023simple,zache2023quantum,su2024cold}. The noisy nature of current quantum simulation technology demands error mitigation/correction techniques to faithfully simulate gauge-invariant dynamics. Over the years, there have been quite a few proposals suggesting different ways to control this error \cite{tran2021faster,halimeh2020reliability,halimeh2021gauge,Halimeh_2022,Schweizer2019,stannigel2014constrained,kasper2023non,stryker2019oracles}. The different proposals and error mitigation schemes led to several digital and analog simulations of Abelian and non-Abelian lattice gauge theories over the last few years \cite{Martinez2016,Kokail_2019,Mil_2020,semeghini2021probing,Yang:2020yer,Atas2021,atas20212,mildenberger2022probing,riechert2022engineering,alam2022primitive,gustafson2022primitive,ciavarella2022preparation,illa2022basic,su2023observation,atas2023simulating,farrell2023scalable,rahman2022self,zhang2023observation,farrell2023preparations,farrell2023preparations2,ciavarella2023quantum,charles2024simulating,mueller2023quantum,kavaki2024square,farrell2024quantum,ciavarella2024quantum}.

In higher dimensions, the Hamiltonian includes plaquette terms, which are gauge invariant but involve the minimum number of sites required to trace a closed path on the lattice. Quantum simulation of the non-local terms poses additional challenges while maintaining non-Abelian gauge invariance locally at each site.
The current work addresses how these complications can be avoided by using the loop- string-hadron (LSH) framework \cite{Raychowdhury:2018tfj, kadam2022loop} towards making concrete progress in quantum simulating QCD. 

The organization of the paper is as follows. In section \ref{sec:lsh}, we briefly review the loop string hadron formalism for SU(3) gauge theory. In section \ref{sec: global}, we focus on the global protection scheme valid for $1+1$ dimensional models. Section \ref{sec: local} discusses the alternate protection scheme, which involves local protection yet is simpler than protecting all the non-Abelian symmetries of the original theory and is valid in higher dimensions. Finally, we summarise in \ref{sec: dis} and discuss the viability of the scheme in terms of its implementation. 


\section{A suitable framework to study dynamics: LSH framework}
\label{sec:lsh}
In this work, we focus on a particular framework, namely the loop-string-hadron (LSH) framework that has been developed for SU(2) \cite{Raychowdhury:2019iki} and  SU(3) \cite{kadam2022loop} gauge theories. The SU(2) framework has been studied extensively and found to be advantageous over other available frameworks to describe the same physics. The key feature of LSH formalism is that it is constructed to be manifestly gauge invariant. The Hilbert space is characterized by on-site quantum numbers, which correspond to gauge invariant and physical loop-string-hadron excitations at each site, and their dynamics are described by a Hamiltonian, which contains combinations of occupation number operator and ladder operators for the associated loop-string-hadron excitations. 

However, encountering non-local excitations is unavoidable in a gauge theory, which is true for the LSH framework. The notion of non-locality in the LSH framework arises from the fact that the gauge invariant configurations defined on a lattice, equivalent to Wilson loops or strings in the theory, are not a tensor product of on-site LSH states belonging to the on-site LSH Hilbert spaces, but instead are local LSH states weaved via Abelian constraints on each link. The LSH Hamiltonian commutes with the Abelian constraint and the spectrum and dynamics match that of the original Kogut-Susskind Hamiltonian for a particular theory. In this work, we emphasize a couple of interesting and useful conclusions regarding using LSH for Hamiltonian simulation that can be applicable and beneficial to use in appropriate setups.

\subsection{LSH basis}

The LSH basis is defined locally at each lattice site which is characterized by a set of integer-valued quantum numbers, namely the loop and string quantum numbers. Loop quantum numbers are bosonic implying the Hilbert space to be of infinite dimension, while the string quantum numbers are fermionic. However, for practical computation one needs to impose a finite cut-off on the bosonic quantum number so that the Hilbert space is of finite dimension.

For SU(2) gauge theory, the LSH basis states at a site $r$ in $1+1$-d is given by $$|n_l,n_i,n_o\rangle_r$$, where $n_l \in \{0,\infty\}$ is the loop quantum number and $n_i,n_o \in \{0,1\}$ are fermionic quantum numbers. Imposing a cut-off $\Lambda$, one would obtain an onsite Hilbert space of dimension $2^2(\Lambda+1)$. 

\begin{figure}[ht]
    \centering
    \includegraphics[width=1\linewidth]{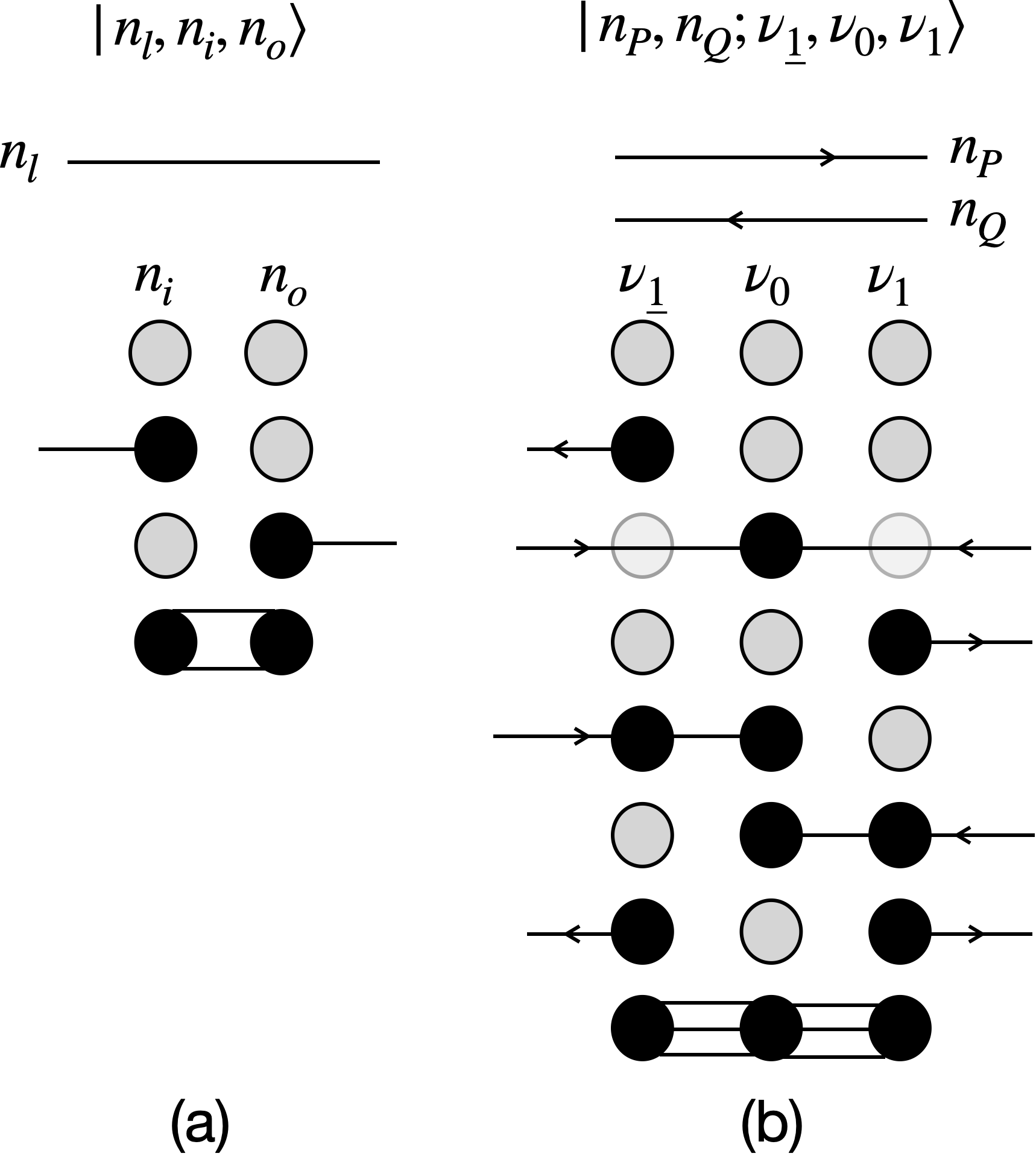}
    \caption{On-site LSH state for (a) SU(2) and (b) SU(3) gauge theories in $1+1$ dimension. Note that the fermionic excitations are denoted as circles, which can be either empty or filled, denoting the corresponding fermionic occupation numbers to be $0$ and $1$, respectively. For SU(2), flux lines are undirected, while for SU(3) being directed, they can flow in two different directions. Within LSH framework, the fermionic occupation numbers are associated with incoming and outgoing strings in different directions. }
    \label{fig-1}
\end{figure}

For the case of SU(3) gauge theory defined on a one-dimensional spatial lattice, the on-site Hilbert space is characterized by two loop and three string quantum numbers as \bea|n_P,n_Q,\nu_{\barone},\nu_0,\nu_1\rangle_r, \label{eq: lsh-state}\eea
where, $n_P,n_Q \in \{0,\infty\}$ and $\nu_{\barone},\nu_0,\nu_1 \in \{0,1\}$. Pictorial representations of the LSH states are given in Fig. \ref{fig-1}. For a bosonic cut-off $\Lambda$, the dimension of the on-site LSH Hilbert space is $2^3(\Lambda+1)^2. $
\subsection{LSH Hamiltonian}
Following the original Kogut-Susskind construction of the Hamiltonian \cite{Kogut:1974ag}, the Hamiltonian in the LSH framework also consists of electric, mass, fermionic hopping, and magnetic field terms. Considering the simplest case of one spatial dimension, the magnetic term is not present, but the Hamiltonian (scaled to make it dimensionless) is written as:
\begin{eqnarray}
    H=H_E+\mu H_m+xH_I  \label{eq:ham}
\end{eqnarray}
where $\mu$ and $x$ are dimensionless parameters that depend upon the lattice spacing, coupling, and mass of the fermions. The strong coupling limit of lattice gauge theory is always interesting as the theory is confined in this regime, defined by $x\rightarrow 0$, keeping $\mu$ to be finite. 
The LSH basis described before is a strong coupling eigenbasis, i.e., in the LSH basis, the matrix representation of $H_E$ and $H_m$ are both diagonal. Also note that, both the terms $H_E$ and $H_m$ are ultra-local in the strong coupling basis, so contributions from all the lattice sites are added to make the electric energy and mass of the entire system. 

In the LSH framework, one can define on-site occupation number operators for the LSH basis both for the SU(2) and SU(3) case, and the onsite contribution of electric and mass term is obtained as a function of occupation numbers of the LSH excitations at that site:
\begin{eqnarray}\hat n_l|\{n_l\}, \{n_f\}\rangle  &=& n_l|\{n_l\}, \{n_f\}\rangle  \label{eq:num1}\\
\hat n_f|\{n_l\}, \{n_f\}\rangle  &=& n_f|\{n_l\}, \{n_f\}\rangle  \label{eq:num2}
\end{eqnarray}
where $\{n_l\}$ denotes the set of bosonic quantum numbers, i.e. $n_l$ for SU(2) and $n_P~\& ~n_Q$ for SU(3). Similarly, 
The LSH Hamiltonian is given in Appendix \ref{app:ham} both for SU(2) and SU(3). 
Interestingly, the mass part of Hamiltonian is effectively proportional to the total occupation number of fermions at each site. The electric part, on the other hand, counts the flux lines that originate or enter a site. A closer inspection of Fig. \ref{fig-1} reveals that non-zero occupation of the fermionic LSH modes often comes with flux lines attached to it, and hence, the fermionic LSH excitations also contribute to the on-site electric energy. The detailed expressions are given in the Appendix \ref{app:ham}.

For a $1+1$ dimensional theory, the only term responsible for dynamics is the hopping or `matter-gauge interaction' term $H_I$. The construction of the interaction Hamiltonian $H_I$ is nontrivial, compared to its SU(2) counterpart as evident in the detail construction given in \cite{kadam2022loop}. This is due to the fact that matter content at a SU(3) site contains much more detailed inner structure than just being present as string end operators for the SU(2) case. This can be qualitatively appreciated from FIG. \ref{fig-1}. To finalize a representation of $H_I$ in LSH basis, we consider the following set of normalized ladder operators, which acting on a on-site LSH state, changes its excitations by $\pm 1$ unit. 
\begin{eqnarray}
   \Gamma^{\dagger}_{L}|\{n_l\}, \{n_f\}\rangle &=& |\{n_l|n_L+1\}, \{n_f\}\rangle \label{eq:lad1} \\
    \chi^{\dagger}_{F}|\{n_l\}, \{n_f\}\rangle &=& (1-\delta_{1,n_F})|\{n_l\}, \{n_f|n_F+1\}\rangle \\
     \Gamma^{}_{L}|\{n_l\}, \{n_f\}\rangle &=& |\{n_l|n_L-1\}, \{n_f\}\rangle \\
    \chi^{}_{F}|\{n_l\}, \{n_f\}\rangle &=& (1-\delta_{0,n_F})|\{n_l\}, \{n_f|n_F-1\}\rangle  \label{eq:lad4}
\end{eqnarray}
The notation $\{n_l|n_L\pm1\}$ and $\{n_f|n_F\pm 1\}$ within the ket denotes that only the specific bosonic $(L)$ or fermionic $(F)$ LSH quantum number has changed, while keeping all other in the set fixed. 

The following definition captures the inner structure of gauge invariant dynamics for this theory:
\begin{widetext}

\begin{eqnarray}
    H_I &=& \sum_r \Bigg[ \sum_{F=\barone,0,1} \chi_F^\dagger (r) \Big[\hat O_1^{(F)}(\{\hat n_l(r),\hat n_f(r)\} ) 
    \times \hat O_2^{(F)}(\{\hat n_l(r+1),\hat n_f(r+1)\} ) \Big]\chi_F(r+1)\Bigg] 
    + \mathrm{h.c. } \label{eq:hi}
\end{eqnarray}
    
\end{widetext}
The exact structure of $\hat O_{1/2}^F(\{\hat n_l(r),\hat n_f(r)\} )$ for $F=\barone,0,1$ are listed in the Appendix \ref{app:ham}, equations (\ref{eq:HI1})-(\ref{eq:HI3}).

\subsection{Gauss law constraints in LSH framework}
The physical implication of gauge invariance is in some non-local structures, and the same is reflected in the non-local Wilson loop and string basis as the gauge invariant basis of gauge theories. However, as listed in FIG, the LSH framework is an intermediate step in building on-site gauge invariant Hilbert space. \ref{fig-1}. Connecting the on-site LSH Hilbert space to the space of Wilson loops and strings requires continuity of flux lines (for each type of directed flux lines for SU(3)) across neighboring lattice sites. This is true for arbitrary dimensions as well.  

Analyzing the $1+1$-dimensional SU(3) theory, let us focus on-site $r$, which is connected to links along direction $1$ and $\barone$. Careful inspection of directed flux lines in FIG. \ref{fig-1} yields:
\begin{eqnarray}
    \label{eq:Pout}
    P_1(r) &=& n_P(r)+ \nu_1(r)\Big(1-\nu_0(r)\Big)  \\
    \label{eq:Qout}
    Q_1(r) &=& n_Q(r)+ \nu_0(r)\Big(1-\nu_{\barone}(r)\Big)
\end{eqnarray}
Where $P_1(r)$ (and $Q_1(r)$ ) denote the number of flux lines going from left to right (and  right to left) respectively on the link along $1$ direction connected to site $r$. 

The same on the link in direction $\barone$ connected to site $r$ is given by:
\begin{eqnarray}
    \label{eq:Pin}
    P_{\barone}(r) &=& n_P(r)+ \nu_0(r)\Big(1-\nu_1(r)\Big)  \\
    \label{eq:Qin}
    Q_{\barone}(r) &=& n_Q(r)+ \nu_{\barone}(r)\Big(1-\nu_{0}(r)\Big)
\end{eqnarray}

Continuity of flux lines or the local LSH states correspond to Wilson loop and string states of the gauge theory demands for the constraints:
\bea
    \label{eq:AGL1}
    P_1(r)= P_{\barone}(r+1)\\
    \label{eq:AGL2}
    Q_1(r)= Q_{\barone}(r+1)
\eea
This implies that the space of Wilson loops and strings are not a direct product of LSH states but rather a projected subspace that satisfies a pair of local Abelian Gauss-law (AGL) constraints given in (\ref{eq:AGL1},\ref{eq:AGL2}).

\subsection{Global symmetries of the LSH framework in $1+1$ dimension}

For a SU(3) gauge theory on one-dimensional staggered lattice, the global symmetry remains to be SU(3) which can be generated by $8$ charges:
\begin{eqnarray}
    Q^a &=& \sum_{r=0,2,..} \Psi^\dagger_\alpha (r) \left(\frac{\lambda^a}{2}\right)^{\alpha}_\beta \Psi^\beta (r) \nonumber \\
    && - \sum_{r=1,3,..} \Psi_\alpha (r) \left(\frac{\lambda^{*a}}{2}\right)^{\alpha}_\beta \Psi^{\dagger\beta} (r)
\end{eqnarray}
where, $\Psi_\alpha (r)$ are fermionic triplets (anti-triplets) (fundamental irreps of SU(3)) located at even (odd) sites $r$ and $\lambda^a$ are Gell-Mann matrices for $a=1,2,...,8$. A global state defined on the lattice can thus be characterized by global SU(3) casimirs $|p,q, I, M, Y\rangle$ of an SU(3) irrep using the standard notation, where $p,q$ is analogous to $j$ quantum number (Casimir) of SU(2) irreps and the set $I,M,Y$ denote the magnetic quantum numbers analogous to $m$ of an SU(2) irrep denoted as $|j,m\rangle$. Naturally, analyzing the global symmetries of the SU(3) gauge theory is more involved than the SU(2) theory when using the irrep-basis or Kogut-Susskind formalism of the theory. However, global symmetries play a crucial role by providing a block diagonal structure of the Hamiltonian and analyzing the entanglement properties of gauge theories, which is important in the context of quantum computation. The global symmetry structure of the same theory in the LSH framework is found to be much more intuitive and can be used for more efficient computations at ease.

The LSH framework, being a solution to all the non-Abelian constraints, is still constrained by the local Abelian Gauss law.  As a consequence, the global symmetries for LSH also become Abelian. For $1+1$-d theory coupled to staggered matter fields, as depicted  in (\ref{eq:hi}), the interaction term manifestly preserves the global occupation numbers of the  three individual gauge invariant fermionic or string modes: 
\begin{eqnarray}
    q_f=\sum_r \nu_{f}(r) \label{eq:global}
\end{eqnarray}
for $f= \barone, 0, 1$.
Alternately, one can also define the linear combinations of $q_{\barone}, q_0, q_1$ as \cite{kadam2022loop}:
\begin{eqnarray}
    \mathcal F &=&  q_{\barone}+ q_0+ q_1 \nonumber \\
    \mathcal P &=& q_1-q_0 \nonumber \\
    \mathcal Q &=& q_0-q_{\barone} \nonumber \\
\end{eqnarray}
Note that $ \mathcal F$ gives the naive total fermionic occupation number for the system, while $ \mathcal P$ and $ \mathcal Q$ give the imbalance between the incoming and outgoing flux of the lattice individually for leftward and rightward flux lines. The Hamiltonian matrix written in the LSH basis is block diagonalized, where each block is characterized by unique set of global quantum numbers $(\mathcal F, \mathcal P, \mathcal Q)$. An ideal Hamiltonian simulation is expected to obey the block diagonal structure of the theory.

A definite benefit of using the LSH framework is the link between these global symmetries and the remnant local Abelian constraints in 1+1 dimension. As illustrated in the case of SU(2) in \cite{mathew2022protecting}, the particular structure of the interaction Hamiltonian for a $1+1$d theory with manifest global symmetries is a consequence of the interaction Hamiltonian being Abelian gauge invariant as well. However, a similar connection is nontrivial to demonstrate for the case of SU(3) via analytic expression as the string end operators are now of three types and involve an intermediate or $0^{th}$ mode rather than only corresponding to incoming and outgoing string. However, the current work establishes the connection between AGL and global symmetries for SU(3), demonstrated numerically in the next section. 

We first consider situations where the bit-flip or other errors present in quantum hardware can lead to deviation from this ideal gauge invariant LSH dynamics. Using two different prescriptions, we provide solutions for gauge protection and numerically establish their validity. These solutions are useful in tensor network simulation, analog simulation protocols, and digital quantum circuits. 

\section{Symmetry protection in erroneous dynamics }
\begin{figure*}
    \centering
    \includegraphics[width=\textwidth]{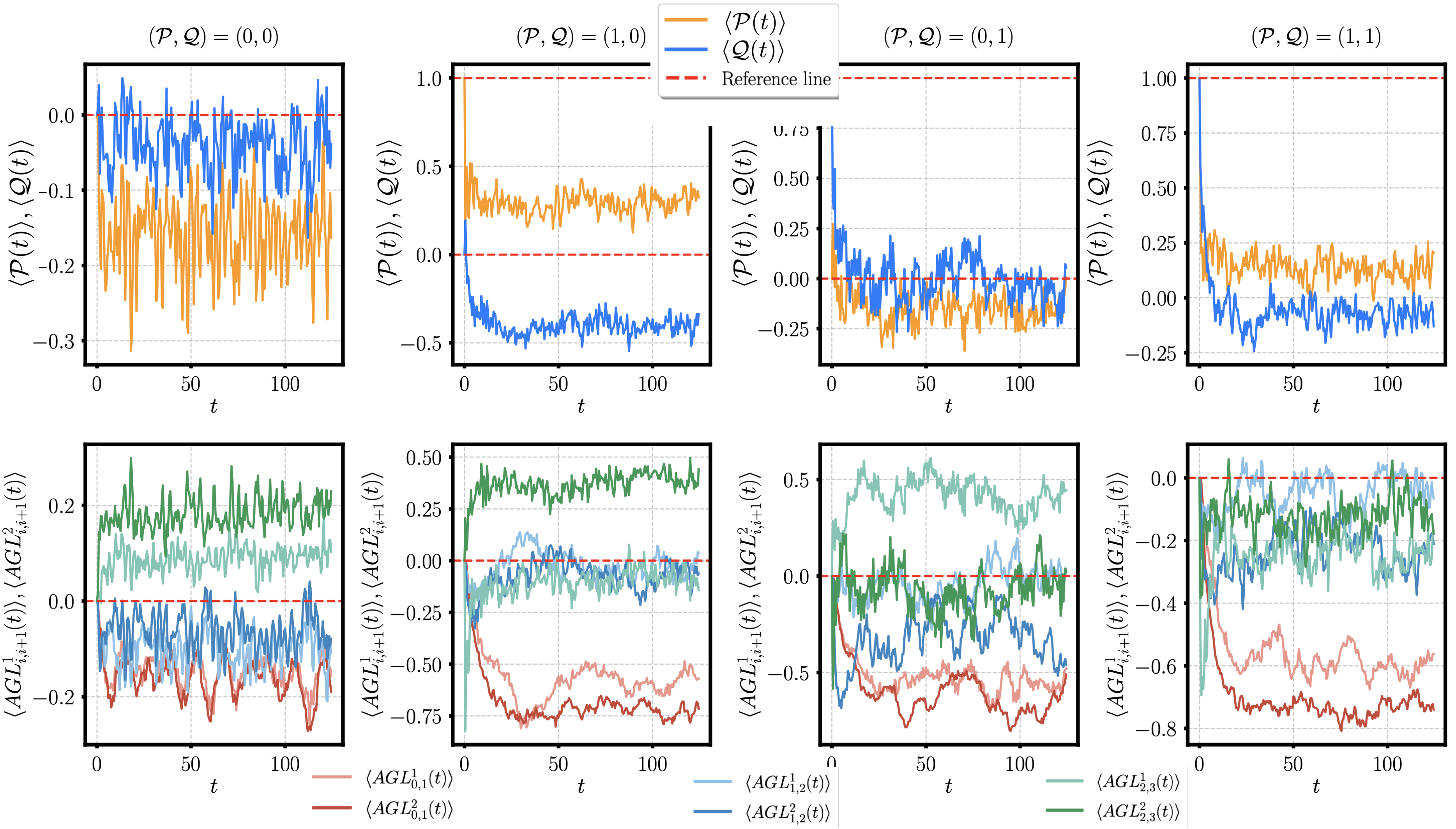}
    \caption{The first row contains the dynamics of expectation values of the global charges $\mathcal P$ and $\mathcal Q$  and the second row contains the same for the six AGLs across three links evolving under the (erroneous) Hamiltonian given in equation (\ref{eq:Herr}). Each column corresponds to the evolution of an initial state in a particular global symmetry sector denoted by $(\mathcal P, \mathcal Q)$. Here, $\langle AGL^1 \rangle_{r,r+1},\langle AGL^2 \rangle_{r,r+1}$ corresponds to equation (\ref{eq:AGL1}),(\ref{eq:AGL2}) respectively. The plots demonstrate that none of the global and local symmetries are preserved by (\ref{eq:Herr}) as the dynamics are not coinciding with reference lines (red dotted lines) for each.}
    \label{fig-violation}
\end{figure*}

We first consider a model where the global symmetries mentioned in (\ref{sec: global}) are violated. Note that, the first global symmetry characterized by $\mathcal F$, is easy to preserve, as that refers to only a perfect state preparation with a fixed total number of qubits given by $\mathcal F$. Even while preserving the value of $\mathcal F$,  the following interaction term violates $\mathcal P, \mathcal Q$, modelling bit-flip error that is most common to occur:
\begin{widetext}

\begin{eqnarray}
    H'_I &=& \sum_r  \sum_{F,F'=\barone,0,1} 
    \Bigg[ (x\delta_{F,F'}+x'(1-\delta_{F,F'})) 
    \Big[\hat O_1^{(F)}(\{\hat n_l(r),\hat n_f(r)\} ) \Big]
    \Big[\hat O_2^{(F')}(\{\hat n_l(r+1),\hat n_f(r+1)\} ) \Big]
    \chi_{F'}(r+1)\Bigg]  \nonumber \\ && + \mathrm{h.c. } \label{eq:hierror}
\end{eqnarray}
    
\end{widetext}
Note that only a set of sub-terms of (\ref{eq:hierror}), i.e. only for $F=F'$ cases in the sum, gives the original interaction Hamiltonian (\ref{eq:hi}). 
One can now write a total Hamiltonian composed of the erroneous interaction as
\bea
\label{eq:Herr}
    H_{err} = H_E + H_M + H'_I
\eea
It is manifest from (\ref{eq:Herr}) that, $H'_I$ violates the global symmetries. We numerically demonstrate in FIG. \ref{fig-violation} that time evolution with the erroneous Hamiltonian is not confined to the AGL satisfying sector of the LSH Hilbert space. 
All our numerical calculations are performed for a 4-staggered site lattice with $x=x'= 1$ to ensure maximal mixing of the three fermionic modes and the mass parameter $\mu=1$. We also impose a cutoff on the Bosonic quantum numbers $n_P,n_Q = 2$. The QuSpin Package \cite{weinberg2017quspin,weinberg2019quspin} is used to construct the Hamiltonian and its subsequent time evolution. The initial state is chosen to be the strong-coupling vacuum (unless otherwise specified), which satisfies the AGLs across all the links and lies in the global symmetry sector $(\mathcal F, \mathcal P,\mathcal Q) = (6,0,0)$. In the LSH quantum numbers given in (\ref{eq: lsh-state}), the initial state is given by :
\bea
\label{str-vac-state}
    \ket{\Psi}_{i} &=& \ket{0,0,0,0,0}_0\otimes\ket{0,0,1,1,1}_1\\\nonumber
    &\otimes&\ket{0,0,0,0,0}_2\otimes\ket{0,0,1,1,1}_3
\eea
The above-defined state is time-evolved using the erroneous Hamiltonian defined in equation (\ref{eq:Herr}) as $\ket{\Psi(t)} = e^{-iH_{err}t}\ket{\Psi}_{i}$. FIG \ref{fig-violation} shows the averages of the AGLs across each link along with the average of the global symmetry operators $\mathcal  P, \mathcal  Q$ as a function of time. The averages move away from the ideal values and signify mixing with non-AGL satisfying states.

\subsection{Scheme I: LSH dynamics via global symmetry protection}
\label{sec: global}
\begin{figure*}[ht]
    \centering
    \includegraphics[width=\textwidth]{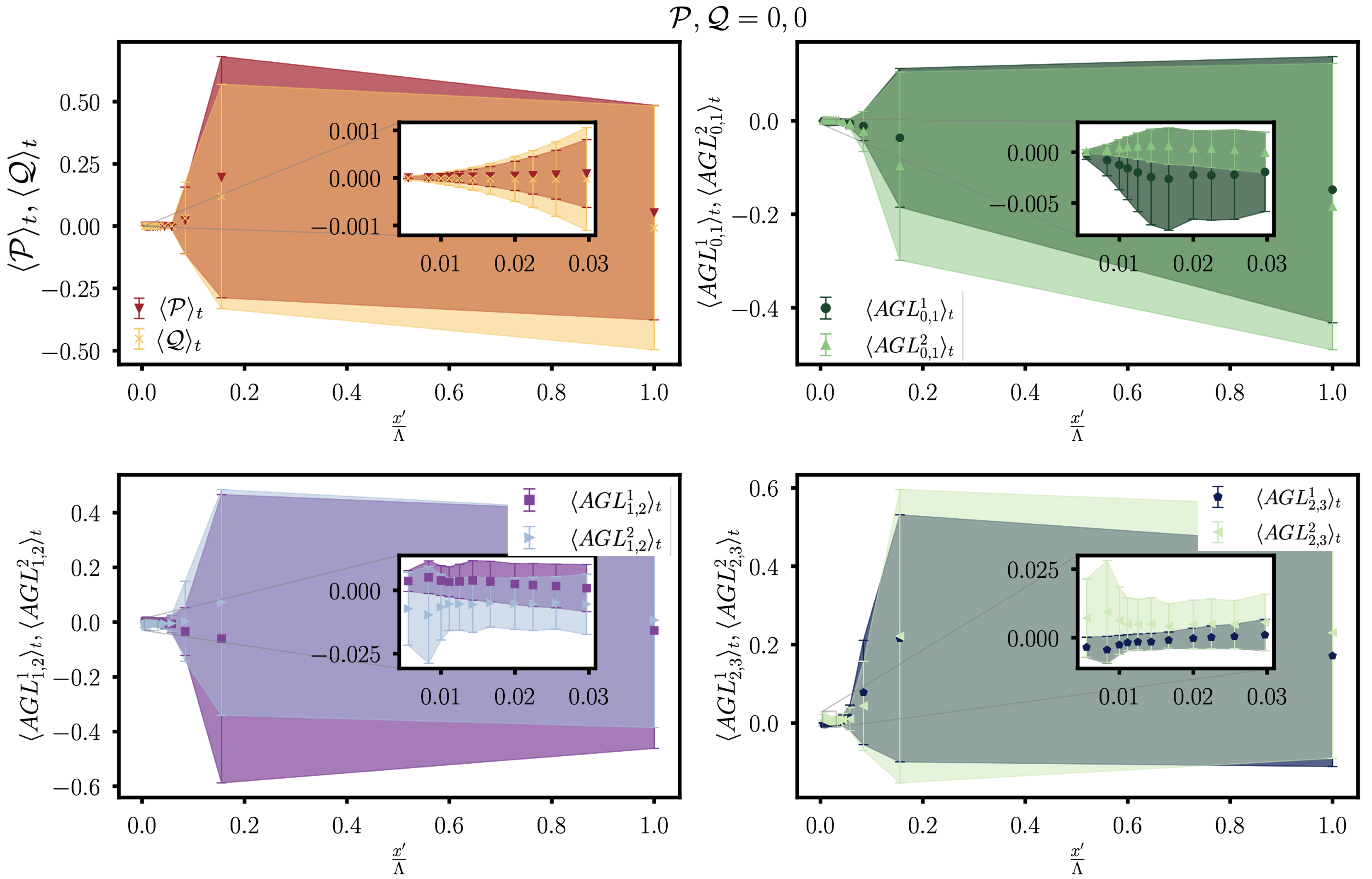}
    \caption{The figure illustrates the temporally averaged values of the global symmetry operators $\mathcal P, \mathcal Q$ and the two Abelian Gauss laws across three links for a 4-staggered site system. For $x=x'=1$, taking the $\Lambda \rightarrow \infty$ limit ensures that $\mathcal P, \mathcal Q$ tend to the sector the initial state belonged to along with the Abelian Gauss laws across the three links converging to zero. The error bars are the temporally averaged operator fluctuations $\mathcal O$ for each $\Lambda$ parameter run}
    \label{fig:protection_global}
\end{figure*}

A scheme for dynamically protecting global symmetries (in other words, projecting the dynamics in different super-selection sectors) of SU(3) gauge theory coupled to dynamical matter fields in $1+1$ dimension generalizing \cite{mathew2022protecting} is presented in this section.
As discussed previously, $H'_I$ violates the super selection rules or the global symmetries of the LSH Hamiltonian.
To counteract this phenomenon, we propose a pair of protection terms that preserve the global symmetries of the SU(3) LSH Hamiltonian, namely the $\mathcal P$ and $\mathcal Q$ operators. We propose adding the following set of operators to the Hamiltonian:
\bea
\label{Hprotect}
    H_{\mathcal P} = \Lambda \Bigg[\mathcal P - \sum_{r=0}^{N-1}\Big[\nu_0-\nu_{\underline{1}}\Big]\Bigg]\\
    H_{\mathcal Q} = \Lambda \Bigg[\mathcal Q - \sum_{r=0}^{N-1}\Big[\nu_1-\nu_{0}\Big]\Bigg]
\eea
Here, $\mathcal P$ and $\mathcal  Q$ can take integer values depending on the symmetry sector of interest. Adding these terms to the total Hamiltonian will ensure that an initially gauge-invariant state can be protected with, at most, a single body protection term. The total Hamiltonian will now read as:
\bea
\label{Hsimulate}
    H = H_E +  H_M + H'_I + H_{\mathcal P} + H_{\mathcal Q}
\eea
where $\Lambda$ is a constant term kept large enough to ensure a sufficient gap exists between the symmetry-protected sector and the other part of the spectrum. The validity of the scheme in confining the dynamics in each super-selection sector is demonstrated in FIG. \ref{fig:protection_global}. 

The numerical study presented in FIG. \ref{fig-violation} demonstrated that the modified interaction Hamiltonian $H'_I$ does, in effect, violate all of the Abelian Gauss laws for the lattice and drive the dynamics into the non-AGL satisfying sector of the LSH Hilbert space. However, the numerical study performed in this work establishes that efficient protection of all of the global symmetries of this theory, in effect, protects each individual Abelian Gauss law originally present in the theory.

Suitable protection strength $\Lambda$ is also capable of ensuring a sufficient gap between the AGL satisfying and AGL violating states in the Hilbert space. Thus, the AGL-violating states are suppressed during time evolution, and the dominant contribution comes from the AGL-satisfying states. As before, the testbed for the numerical demonstrations is a 4-staggered site lattice with $x=x'=1$, $\mu=1$, and the strong coupling vacuum as our initial state defined in equation (\ref{str-vac-state}). 
Time evolving this initial state under the new Hamiltonian defined in equation (\ref{Hsimulate}) will ensure that the AGLs and global symmetries are restored in the limit of $\Lambda \rightarrow \infty$. FIG \ref{fig:protection_global} shows how the local and global symmetries are being restored. The error bars are the quantum fluctuations of the operators of interest, mainly $\{AGL^{\{1,2\}}_{r,r+1},\mathcal P,\mathcal Q\}$.  These operators are temporally averaged for a short time of 0.5T, where $T$ is of the order $10^3$. This is because the size of the Hilbert space for the 4-site lattice is of the order $10^7$, resulting in the norm of the Hamiltonian defined in equation (\ref{Hsimulate}) becoming very large. This renders it difficult to simulate for larger time scales. 
If one prepares an initial state in the correct global symmetry sector that also satisfies the AGLs across each link, the symmetry protection protocol will ensure that the large $\Lambda$ limit will approximately conserve the global symmetries and the local constraints. With the state guaranteed to preserve the global symmetries of the theory, we can compute the dynamics of relevant observables of the theory, namely the staggered fermion density and the electric Hamiltonian density across the lattice, and compare them with the gauge-invariant dynamics. This scheme also applies to initial states belonging to different global symmetry sectors, and Appendix \ref{app:} contains the results for such a choice of initial state.

\subsection{Scheme II: LSH dynamics via local symmetry protection scheme }
\label{sec: local}
As it is manifest that $H'_I$ violates each individual AGL, the most straightforward way to protect the AGLs is by adding protection terms to the erroneous Hamiltonian corresponding to each. 
In this section, we demonstrate the same to be working for the erroneous Hamiltonian given in (\ref{eq:Herr}). 
 In contrast with the previous section, where we used a single body protection term to simulate an approximate gauge theory, here the focus is on two body protection terms which is just equivalent to imposing the square of the two AGLs across each link. As before, we consider equation (\ref{eq:hierror}) to be the erroneous interaction term.

The full Hamiltonian will now modified to include the following two terms:
\bea
    \label{eq:AGLprot1}
    H_P
    = \Lambda  \sum_{r=0}^{N-1} \Bigg[P_{1}(r) - P_{\underline{1}}(r+1)\Bigg]^2\\
    \label{eq:AGLprot2}
    H_Q
    = \Lambda  \sum_{r=0}^{N-1} \Bigg[Q_{1}(r) - Q_{\underline{1}}(r+1)\Bigg]^2
\eea
where $P_1,P_{\underline{1}},Q_1,Q_{\underline{1}}$ are defined according to equation (\ref{eq:Pout}-\ref{eq:Qin}).
The newly modified Hamiltonian can now be written out as follows:
\bea
\label{eq: local-protection}
    H = H_E + H_M + H'_I + H_P
    + H_Q
\eea

\begin{figure*}
    \centering
    \includegraphics[width=\textwidth]{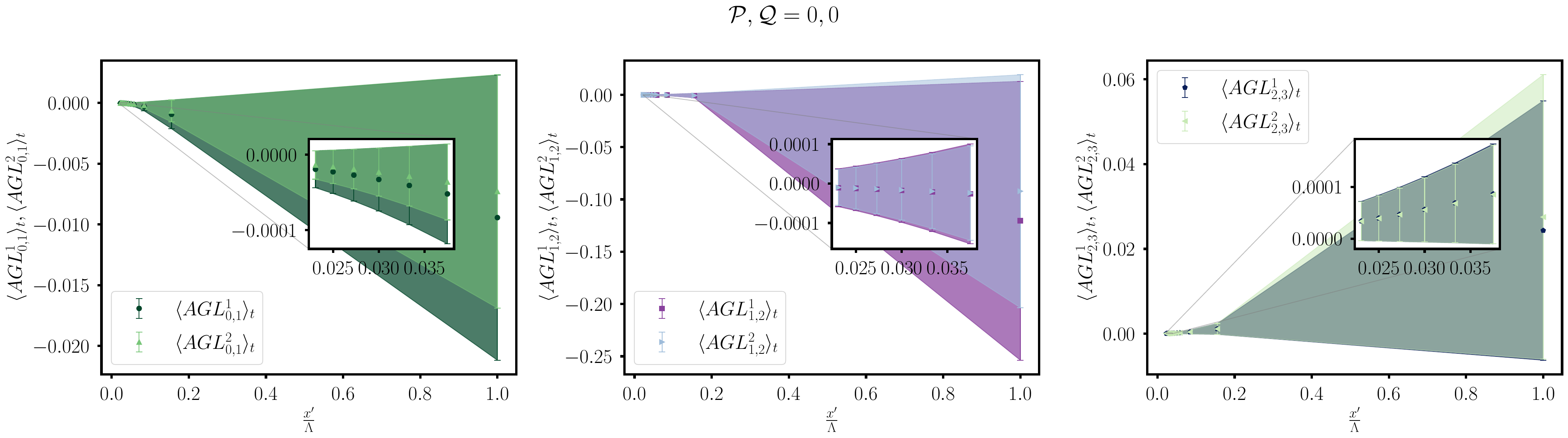}
    \caption{Operator average of the local AGLs and the global symmetry operators as a function of increasing $\Lambda$. The error bars correspond to the temporal average of the operator fluctuation.}
    \label{fig:local_sym}
\end{figure*}

Once again, we repeat the numerical experiment of starting with the state defined in equation (\ref{str-vac-state}) and measure the AGLs as a function on $\Lambda \rightarrow \infty$. In FIG. \ref{fig:local_sym}, we demonstrate that this scheme with a suitable choice of protection strength $\Lambda$ allows the dynamics to remain confined into the AGL satisfying sector of the Hilbert space.  
The $\Lambda \rightarrow \infty$ limit is where an approximate gauge theory is expected to be recovered, and as per the numerical simulation, we see that the operator averages tend to the desired values, with the expectation value of the AGL to be vanishing.    Additionally, the temporally average operator fluctuations are also of the order $10^{-5}$, indicating that the non-AGL satisfying states' contributions are becoming negligible.

Such a local Abelian protection can be applied to higher dimensions where it is indeed essential. In this work, we do not present calculations for higher dimensions as that become computationally heavy and the one-dimensional study is sufficient to demonstrate the proof of principle.

\begin{figure}[h]
    \centering
    \includegraphics[width=0.5\textwidth]{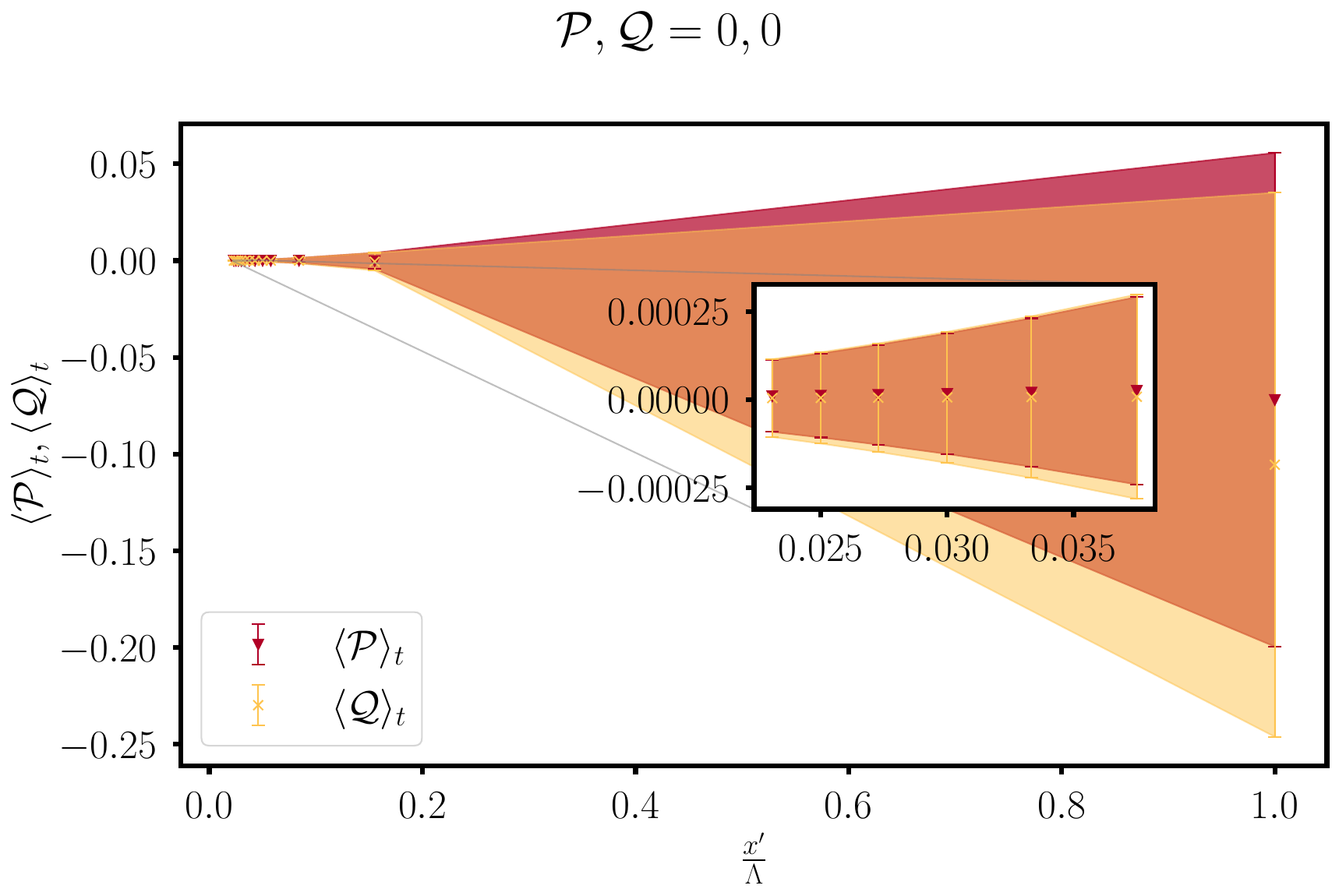}
    \caption{Super-selection rules are preserved by protecting the local symmetries. The initial state is chosen in the $0,0$ sector and evolved with (\ref{eq: local-protection}). }
    \label{fig:local_PQ}
\end{figure}

The bonus finding in this one-dimensional demonstration of the local protection scheme is that each successful protection of local symmetry also guarantees that the dynamics are confined to the original global symmetry sectors (as per the initial state prepared) of the LSH Hamiltonian. This is also demonstrated in FIG. \ref{fig:local_PQ}. 

\section{Conclusions and outlook}
\label{sec: dis}
The current work demonstrates liabilities of a simple scheme of protecting all the symmetries of non-abelian gauge theories in simulating its dynamics. The major challenge in simulating non-Abelian gauge theories is two fold. The first non-trivial task is preparing the initial state which is global state, yet gauge singlet at each and every lattice site. As the Hamiltonian is expected to commute with the generators of all gauge transformations, the dynamics ideally should remain confined in a subspace of the entire gauge theory Hilbert space. However, in terms of erroneous quantum devices, this is not expected to be the reality and hence the simulation scheme demands for a symmetry protection protocol, for any available platform. Following the warm-up exercises towards quantum simulating simpler gauge theories carried out over more than a decade now, the community is now focussing on the SU(3) gauge theories \cite{ciavarella2021trailhead,PhysRevResearch.5.033184, ciavarella2023quantum, ciavarella2024quantum} in $1+1$ dimension and the final aim is to perform simulations in higher dimension. The major restriction in achieving the goal is definitely imposing the complicated SU(3) symmetries and protecting the same. The loop-string-hadron framework for SU(3) in \cite{kadam2022loop} provides a convenient framework for Hamiltonian simulation, where the basis states are inherently SU(3) singlet, leading to no need of imposing any SU(3) symmetry at any point of calculating or simulating synamics. The current protocol, takes advantage of this fact and provides a viable scheme for symmetry protection demonstrating the advantage of using LSH framework.

$1+1$ dimensional gauge theories, although offer the first test-bed for novel simulation protocols are often a lot easier to analyze as the gauge field is not dynamic. However, the global symmetry still possesses a rich non-Abelian structure implying the superselection rules to be governed by SU(3) symmetry and its representations. The LSH framework, being fully Abelianized leads to simpler Abelian global symmetry sectors which is far more intuitive. The current protocol utilizes the same for protecting all the symmetries in $1+1$ dimensions. This is the major advantage of enabling one to simulate SU(3) gauge theory without imposing either any non-abelian nor any local symmetries. Work is in progress in utilizing this scheme for developing analog quantum simulation protocols, which will be reported shortly. This scheme also remains useful for multiple directions of ongoing research efforts in developing tensor network algorithms for calculating dynamics \cite{banuls2023tensor, PhysRevLett.132.091903, knaute2024entanglement, feldman2024superselection}, as well as quantum simulation of scattering of wave-packets \cite{davoudi2024scattering, bennewitz2024simulating}, or studying thermalization dynamics of gauge theories \cite{ebner2024eigenstate, davoudi2023towards,mueller2023quantum} using quantum simulators. Works are progressing towards generalizing these schemes to deliver building blocks for the ultimate goal of quantum-simulating QCD, and this current work is one of those.

\section*{Acknowledgements}
 We acknowledge discussions with Pubasha Shome. Research of IR is supported by the  OPERA award (FR/SCM/11-Dec-2020/PHY) from BITS-Pilani, the Start-up Research Grant (SRG/2022/000972) and Core-Research Grant (CRG/2022/007312) from SERB, India and the cross-discipline research fund (C1/23/185) from BITS Pilani. 

\bibliography{main}
\appendix

\section{LSH SU(3) Dictionary}
\label{app:ham}
The LSH formulation for the SU(3) gauge group follows a similair albeit involved construction\cite{kadam2022loop}. The salient features are neverthess identical to that of the SU(2) formulation. Here, we briefly review the key ideas necessary to understand the construction of the basis states and the Hamiltonian. On a 1+1D lattice, the SU(3) LSH Hilbert space is characterized by five quantum numbers $\ketsu$. Here, $n_P$ and $n_Q$ are bosonic quantum numbers which take values from $\{0,\infty\}$. These bosonic quantum numbers represent gluonic fluxes that are oriented in either the ($\underline{1}\rightarrow 1$) direction or ($1\rightarrow \underline{1}$). The fermionic degrees of freedom are characterized by $\nu_{\underline{1}},\nu_0,\nu_1 \in \{0,1\}$. Since SU(3) is a rank-2 group, the irreducible representations (irreps) are characterized by two indicies, denoted as $P$ and $Q$ (both taking positive integer values). At a particular site, one can associate irreps to both sides of the site such that $P(\underline{1},r),Q(\underline{1},r)$ denotes the indicies to the $\underline{1}$ side of site r and $P(1,r),Q(1,r)$ denotes the indices to the $1$ side of site r. The $P,Q$ integers are related to the SU(3) LSH quantum numbers in the following manner:
\begin{eqnarray}
    P(\underline{1},r) = n_P(r) + \nu_0(r)(1-\nu_1(r))\\
    Q(\underline{1},r) = n_Q(r) + \nu_{\underline{1}}(r)(1-\nu_0(r))\\
    P(1,r) = n_P(r) + \nu_1(r)(1-\nu_0(r))\\
    Q(1,r) = n_Q(r) + \nu_0(r)(1-\nu_{\underline{1}}(r))
\end{eqnarray}
The operator constructions are such that they are locally SU(3) invariant and as before, we have local Abelian Gauss laws. The AGL's for the SU(3) group are as follows (expanded form of (\ref{eq:AGL1},\ref{eq:AGL2})):
\begin{eqnarray}
    n_P + \nu_1(1-\nu_0)\Big\vert_r = n_P + \nu_0(1-\nu_1)\Big\vert_{r+1}\\
    n_Q + \nu_0(1-\nu_{\underline{1}})\Big\vert_r = n_Q + \nu_{\underline{1}}(1-\nu_0)\Big\vert_{r+1}
\end{eqnarray}
The AGLs will be used to weave together the gauge-invariant basis states, as was done for the SU(2) construction.
The relevent operator dictionary is provided below. The diagonal operators are defined as follows:
\begin{align}
    \hat{n}_P\ketsu &= n_P\ketsu\\
    \hat{n}_Q\ketsu &= n_Q\ketsu\\
    \hat{\nu}_{\underline{1}}\ketsu &= \nu_{\underline{1}} \ketsu\\
    \hat{\nu}_0\ketsu &= \nu_0\ketsu\\
    \hat{\nu}_1\ketsu &= \nu_1\ketsu
\end{align}
The Hamiltonian contains operators which are functions of the above set of operators but their action on the basis states are trivially defined according to the definitions above.
The bosonic ladder operators can be defined as follows:
\begin{align}
    \hat{\Gamma}^{\pm}_P \vert n_P,n_Q;\nu_{\underline{1}},\nu_0,\nu_1 \rangle &=& \vert n_P\pm 1,n_Q;\nu_{\underline{1}},\nu_0,\nu_1\rangle\\
    \hat{\Gamma}^{\pm}_Q \vert n_P,n_Q;\nu_{\underline{1}},\nu_0,\nu_1 \rangle &=& \vert n_P,n_Q\pm 1;\nu_{\underline{1}},\nu_0,\nu_1\rangle
\end{align}
while the fermionic ones are defined as:
\begin{align}
    \hat{\chi}_{\underline{1}}\ketsu &= (1-\delta_{0,\nu_{\underline{1}}})\ket{n_P,n_Q;\nu_{\underline{1}}-1,\nu_0,\nu_1}\\
    \hat{\chi}^{\dagger}_{\underline{1}}\ketsu &= (1-\delta_{1,\nu_{\underline{1}}})\ket{n_P,n_Q;\nu_{\underline{1}}+1,\nu_0,\nu_1}\\
    \hat{\chi}_{0}\ketsu &= (1-\delta_{0,\nu_0})\ket{n_P,n_Q;\nu_{\underline{1}},\nu_0-1,\nu_1}\\
    \hat{\chi}^{\dagger}_{0}\ketsu &= (1-\delta_{1,\nu_0})\ket{n_P,n_Q;\nu_{\underline{1}},\nu_0-1,\nu_1}\\
    \hat{\chi}_{1}\ketsu &= (1-\delta_{0,\nu_0})\ket{n_P,n_Q;\nu_{\underline{1}},\nu_0,\nu_1-1}\\
    \hat{\chi}^{\dagger}_{1}\ketsu &= (1-\delta_{1,\nu_0})\ket{n_P,n_Q;\nu_{\underline{1}},\nu_0+1,\nu_1}
\end{align}
The fermionic operators $\hat{\chi}_{f}$, where $f \in \{\underline{1},0,1\}$, satisfy the canonical anti-commutation relations:
\begin{eqnarray}
    \acomm\Big{\hat{\chi}_f(r)}{\hat{\chi}_{f'}(r')} = \acomm{\hat{\chi}^{\dagger}_f(r)}{\hat{\chi}^{\dagger}_{f'}(r')} = 0\\
    \acomm\Big{\hat{\chi}_f(r)}{\hat{\chi}_{f'}(r')} = \delta_{f,f'}\delta_{r,r'}
\end{eqnarray}
With the Hilbert space and the action of the relevent operators on it explicitly stated, we now define the SU(3) LSH Hamiltonian as follows:
\begin{eqnarray}
    H^{SU(3)}_{LSH} = H_E + H_M + H_I
\end{eqnarray}
The electric and mass Hamiltonian are diagonal in the LSH basis and are defined as
\begin{align}
    H_M &= \mu \sum_r (-1)^r (\hat{\nu}_{\underline{1}}(r)+\hat{\nu}_0(r)+\hat{\nu}_1(r))\\
    H_E &= \sum_r \Bigg[\frac{1}{3}\Bigg(\hat{P}^2(1,r)+\hat{Q}^2(1,r)+\hat{P}(1,r)\hat{Q}(1,r)\Bigg)\\
    &+\hat{P}(1,r)+\hat{Q}(1,r)\Bigg]\nonumber
\end{align}
Finally, the matter-gauge interaction Hamiltonian is of the following form: 
\begin{eqnarray}
    H_I &=& \sum_{r=0}^{N-1} \Bigg[ \sum_{F=\barone,0,1} \chi_F^\dagger (r) \Big[\hat O_1^{(F)}(\{\hat n_l(r),\hat n_f(r)\} ) \nonumber \\
    &&\times \hat O_2^{(F)}(\{\hat n_l(r+1),\hat n_f(r+1)\} ) \Big]\chi_F(r+1)\Bigg] \nonumber \\
    && + \mathrm{h.c. }
\end{eqnarray}
where, the operators $O^F_{1,2}$ are defined as
\begin{widetext}

\begin{align}
\label{eq:HI2}
O^{(\underline{1})}_1 &= {\left[\left(\hat{\Gamma}_Q\right)^{1-\hat{\nu}_0} \sqrt{1+\hat{\nu}_0 /\left(\hat{n}_Q+1\right)} \sqrt{1+\hat{\nu}_1 /\left(\hat{n}_P+\hat{n}_Q+2\right)}\right]} \\
O^{(\underline{1})}_2 &= \left[\sqrt{1-\hat{\nu}_0 /\left(\hat{n}_Q+2\right)} \sqrt{1-\hat{\nu}_1 /\left(\hat{n}_P+\hat{n}_Q+3\right)} \left(\hat{\Gamma}_Q\right)^{\hat{\nu}_0}\right]
\end{align}

\begin{align}
\label{eq:HI3}
O^{(0)}_1 &= \left.\left[\left(\hat{\Gamma}_P\right)^{1-\hat{\nu}_1}\left(\hat{\Gamma}_Q^{\dagger}\right)^{\hat{\nu}_1} \sqrt{1+\hat{\nu}_1 /\left(\hat{n}_P+1\right.}\right) \sqrt{1-\hat{\nu}_{\underline{1}} /\left(\hat{n}_Q+2\right)}\right] \\
O^{(0)}_2 &= \left.\left[\sqrt{1-\hat{\nu}_1 /\left(\hat{n}_P+2\right.}\right) \sqrt{1+\hat{\nu}_{\underline{1}} /\left(\hat{n}_Q+1\right)} \hat{\chi}_0\left(\hat{\Gamma}_P\right)^{\hat{\nu}_1}\left(\hat{\Gamma}_Q^{\dagger}\right)^{1-\hat{\nu}_1}\right]
\end{align}

\begin{align}
\label{eq:HI1}
O^{(1)}_1 &= \left[\left(\hat{\Gamma}_P^{\dagger}\right)^{\hat{\nu}_0} \sqrt{1-\hat{\nu}_0 /\left(\hat{n}_P+2\right)} \sqrt{1-\hat{\nu}_{\underline{1}} /\left(\hat{n}_P+\hat{n}_Q+3\right)}\right] \\
O^{(1)}_2 &= \left[\sqrt{1+\hat{\nu}_0 /\left(\hat{n}_P+1\right)} \sqrt{1+\hat{\nu}_{\underline{1}} /\left(\hat{n}_P+\hat{n}_Q+2\right)} \left(\hat{\Gamma}_P^{\dagger}\right)^{1-\hat{\nu}_0}\right]
\end{align}
\end{widetext}
\newpage 
\section{Results for different initial-states}
\label{app:}
\begin{widetext}
\begin{figure*}
    \centering
    \includegraphics[width=\textwidth]{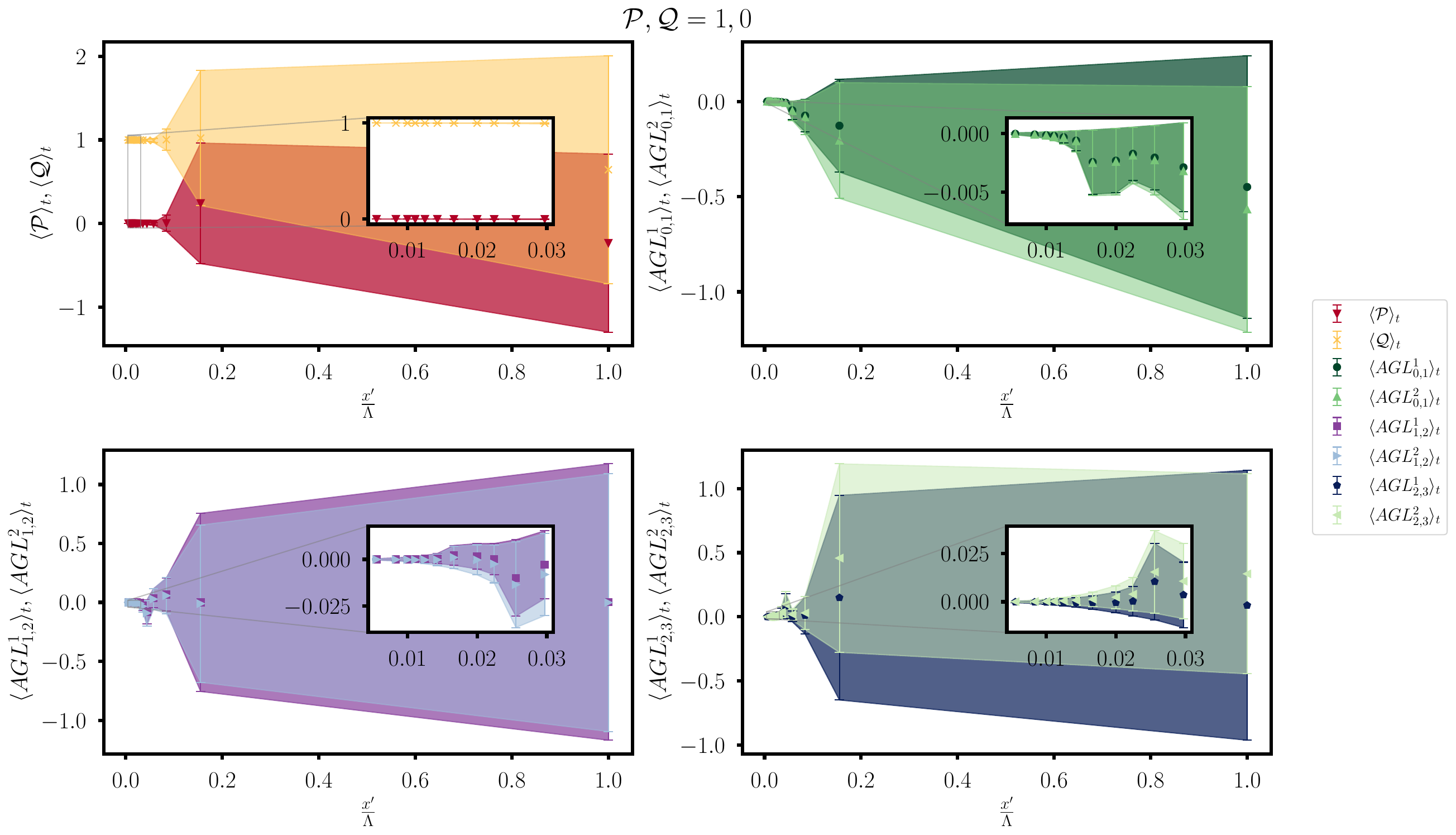}
    \caption{Demonstration of symmetry protection using global scheme for (1,0) sector.}
    \label{fig:protection_global_diffstate}
\end{figure*} 
\end{widetext}

We present the results for an initial state that belongs to a different global symmetry sector.  The initial state is chosen as (belonging to the global symmetry sector $\mathcal{P},\mathcal{Q}=1,0$):
\begin{widetext}
    \bea
\ket{\Psi}_{ini} &= \ket{0, 0, 0, 0, 0}_0 \otimes \ket{0, 0, 0, 0, 0}_1 \otimes \ket{0, 0, 0, 1, 1}_2 \otimes \ket{0, 0, 1, 0, 1}_3 \nonumber
\eea
\end{widetext}
The parameters used for the simulation are the same, with  $x=x'=1$ and $\mu = 1$ for time evolution with the Hamiltonian given in (\ref{Hsimulate}). 
FIG. \ref{fig:protection_global_diffstate} illustrates the temporally averaged values of the global symmetry operators $\mathcal P, \mathcal Q$ and the two Abelian Gauss laws across three links for a 4-staggered site system. Taking the $\Lambda \rightarrow \infty$ limit ensures that $\mathcal P, \mathcal Q$ tends to the sector the initial state belonged to along with the Abelian Gauss laws across the three links converging to zero. The error bars are the temporally averaged operator fluctuations $\mathcal O$ for each $\Lambda$ parameter run. We skip presenting the results of the local protection scheme for other global symmetry sectors, which is also valid. Starting with this particular initial state and evolving with the AGL-protected Hamiltonian as given in (\ref{eq: local-protection}) confines the dynamics in the same super-selection sector. 

\end{document}